\documentclass[aps,pra,twocolumn,showpacs]{revtex4}
\usepackage{graphicx}
\usepackage{dcolumn}
\usepackage{color}
\usepackage{mathrsfs}
\usepackage[breaklinks,colorlinks = true,linkcolor = blue,urlcolor=blue,citecolor=blue]{hyperref}
\usepackage{amsmath}
\usepackage{soul}
\usepackage{lipsum}
\usepackage{amssymb}
\usepackage{epstopdf}
\usepackage{subfigure}

\begin{document}
\title{Transverse-Anderson localization of light in disordered silicon pores}

\author{S. Erfanifam$^{1}$}
\email{s.erfanifam@gmail.com} 
\author{C. Yanik$^{2}$}
\author{M. Erfanifam$^{3}$}

\affiliation{
$^{1}$Shanghai Center for Complex Physics, Department of Physics and Astronomy, Shanghai JiaoTong University, Shanghai 200240, China\\
$^{2}$Sabanci University Nanotechnology Research and Application Center, Tuzla, 34956 Istanbul, Turkey\\
$^{3}$Department of physics, Institute for Advanced Studies in Basic Sciences (IASBS), Zanjan, Iran\\
}
   
\begin{abstract}
Transverse light localization of Anderson type through disorders has made intriguing fundamental topics. Toward this, we found that in the disordered porous Si (PSi) a nearly "zero reflection" (from 450 to 900 nm) can be observed. The localization length versus disorder parameter is experimentally visualized and calculated based on a self consistent theory and solution of stochastic Helmholtz equation for electric fields traveling in a transverse disordered media with the results determined to be in a reasonable agreement with experiments. Based on these calculations for different pore diameters in a constant order parameter, diffusion length can vary from 260 nm to $2.4 \mu m$, accounting as controllability of such structure for further applications in photonic devices. Reflection from ordered PSi structure are evaluated that shows no similar responses to those seen in disordered PSi.
\end{abstract}

\date{\today}
\maketitle
\section{Introduction}
In recent years there has been growing interest in studies of the localization in random media to explain wide range of intriguing complex transport phenomena \cite{arikawa2017,sagev2013,Gaurasundar2014}. Localized states are a kind of reaction to suppress diffusive behavior of the either classical or quantum waves in (bosonic)fermionic systems. The light localization phenomena with superposition of destructively-multiple-scattered quasiparticles is phenomenologically similar to the boson peak in disordered electronic systems where the density of state is rising in certain conditions at low temperatures\cite{schirmacher2006}. In optics, this effect was seen in photonic crystals in both cavity-form arrays made by ion etching techniques and nanoposts as disorder-engineered metasurface \cite{shayan2007,moose2018}. (dis)ordered  pore arrays in silicon as an example of the cavity form photonic crystal can be a promising candidate for future applications in quantum technology, energy harvesting, and biophotonics\cite{crane2017,siu2012}. For these applications controlability of the (dis)order parameter, as a crucial parameter of the Anderson localization effect, is essential. This parameter and its effect on the localization of light in porous silicon, as a cavity form of photonic crystal, has not been investigated. 

In this research we are going to directly visualize the (dis)order parameter and Anderson localization on the optical reflection. We show that multiple scattering of electromagnetic waves with wavelengths higher than the pore diameter in randomly distributed pores can give rise to reduce diffusive propagation of the waves and increases absorption probability of these waves in the PSi. In other words, the interference of scattered waves can renormalize the diffusion coefficient to zero which leads to an energy absorption. In this paper this effect experimentally and theoretically will be investigated. Comparison of our data with the results reported on the optical properties of highly ordered PSi in several reports \cite{brodoceanu2014,niu2016} indicates that randomly distribution of the pores at the same geometry can considerably reduce the light reflection.

The experiment and theory presented here is analogy of electron transport in the disordered photonic media. Theoretical considerations for electron transport based on analytical solution of the stochastic Helmholtz equation have already been thoroughly investigated. In this research we will use the same approach for photons (electromagnetic waves) traveling in our PSi crystal as a (dis)ordered-photonic media.  

\section{Experiment}
\begin{figure}
\begin{center}
\includegraphics[scale=0.42]{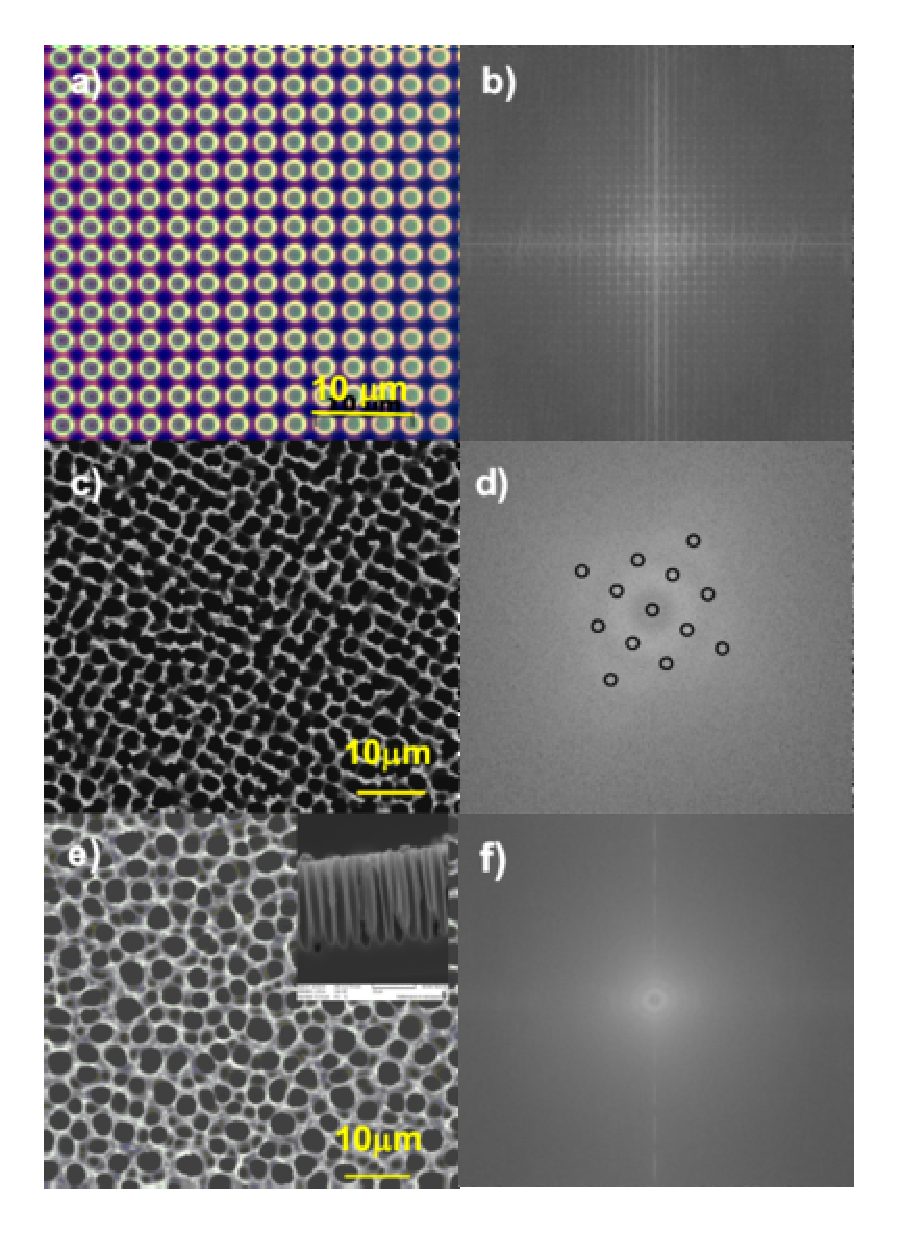}
\end{center}
\vspace{-15pt}
\caption{Optical images, SEM and related FFT analysis of the highly ordered, slightly ordered and disordered samples indicating the order parameter evolution. In contrast to the FFT pattern of the disordered pores in f, ordered pores exhibits some spots in d. The scale of the right panels are adjusted for clarity.}
\label{fft}
\end{figure}

Fabrication of PSi is conducted by anodization (anodic electrochemical etching) process in an appropriate solution which in this case is made of HF and ethanol with 1:4 ratio. The PSi samples investigated in this research are fabricated from n-type silicon (phosphor doped) with 1-10$\Omega$.cm resistance. The anodization process were conducted in a home-made chemical cell in which only one side of the silicon as anode is etched by electrolyte mentioned above to create the pores. The anodization was carried out at room temperature in the presence of graphite/Pt as a cathode. These materials as cathode have low reactivity as well as high resistance against corrosion. Applied current and voltage versus time are controlled. These parameters are important for the uniform formation of the PSi structures. Depending on the required morphology and porosity, either constant current or constant voltage technique was employed. Anodization was conducted under different constant currents/voltages for different anodization times and then analyzed by scanning electron microscopy (SEM) to measure some pore characteristics such as pore height and diameter (see inset of Fig.\ref{fft}e).

In this research we tried to make various sample with different pore-array ordering. Generally, anodization of the bare silicon results in a self-assembled and disordered array of the pores. In order to make ordered pores by selective etching, lithographically  patterned silicons having square arrayed pores with 2$\mu$m pore diameter and 3$\mu$m interpore distance, are employed. 

Figures\ref{fft}a and c show the lithographically patterned silicon before and after anodization, respectively. However, after the pore formation, due to over etching of interpores some deviations from fully ordered array by merging some pores on each other is observed. Finally, the square pattern of the pores is clearly seen. In the Fig.\ref{fft}e SEM micrograph of a sample with fully disordered pores (with comparable pore diameter and pore length values of the ordered one) is exhibited. 

Fast Fourier transform (FFT) taken from images of the photonic-crystals similar to the reciprocal space in condensed matter is commonly used to evaluate the ordering parameter range. The FFT patterns are symmetric about the origin. Right panels in Figs.\ref{fft}  display the calculated FFT patterns at the counterpart left panels. By moving from Fig.\ref{fft}b to d and e, it is clearly seen that bright spots are decreasing and going to be smeared out in fully disordered pores(equivalent to lowering of the pore's ordering range).

\section{Result and discussion}
\begin{figure}[t]
\begin{center}
\includegraphics[scale=0.38]{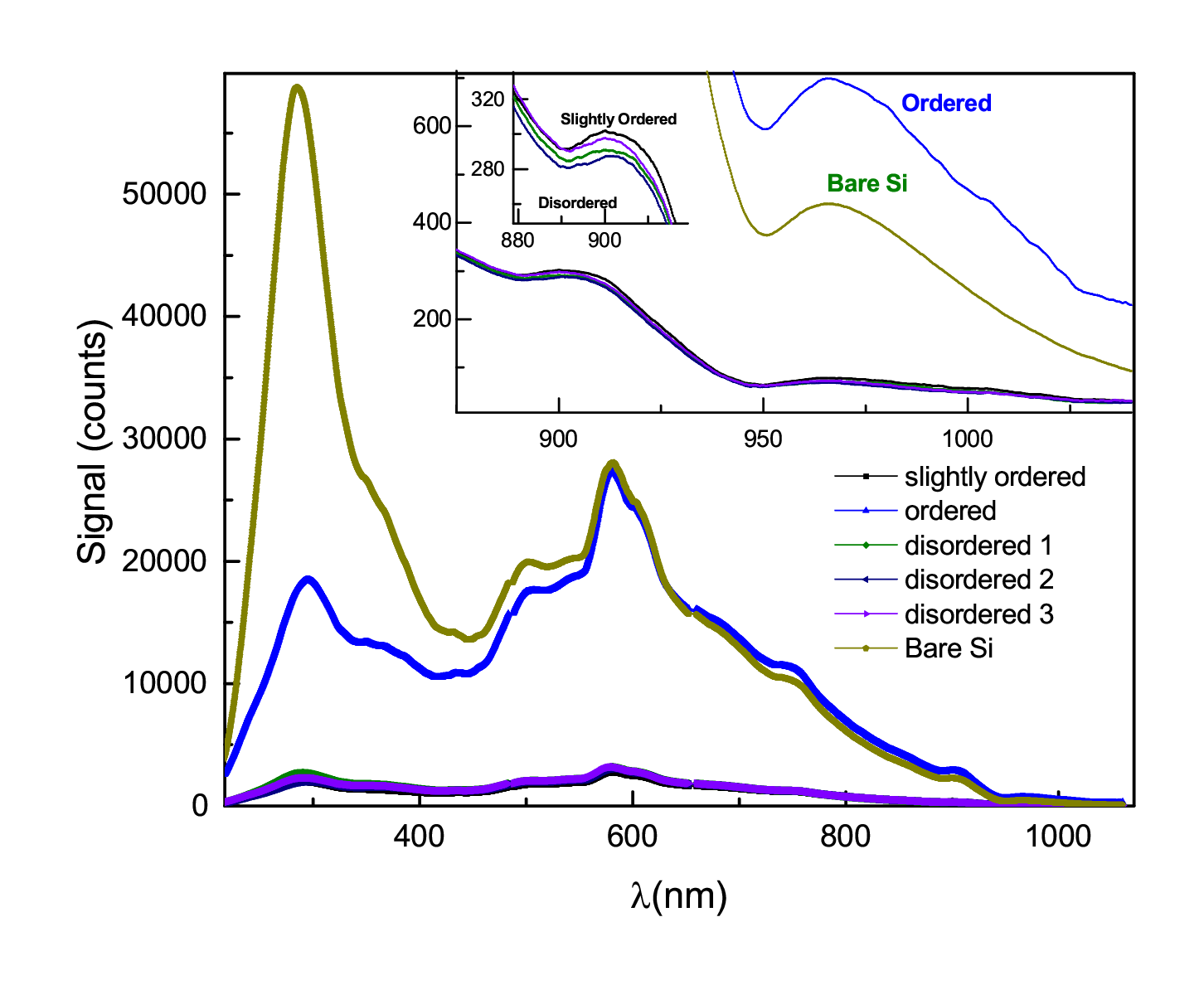}
\end{center}
\vspace{-15pt}
\caption{(Color online)Reflectance spectra of PSi for bare Si, ordered and three disordered PSi (disordered 1,2,3). Inset displays higher range of the wavelength in which Anderson localization is dominant. Inner inset shows increasing of the reflection from slightly ordered sample compared to disordered ones.}
\label{refl}
\end{figure}

In order to study the light-matter interaction, spectroscopy of the reflected light from PSi can be a powerful technique. Fig.\ref{refl} exhibits reflectivity versus wavelength for various samples made on either patterned or nonpatterned silicon. Interestingly, we observed that the reflection from diordered pores is strongly suppressed specially above 600 nm and becomes almost zero by approaching to the 1000 nm. However, below 450 nm at UV region, the reflection is relatively higher. This spectrum has been taken from different disorderd pores with different fabrication parameters. In all cases the reflection spectrum exhibits almost the same behavior. Surprisingly, the reflection from ordered pore arrays is much higher than those disordered ones. It can be seen that the reflection is even higher than the bare silicon where seems to be a kind of light amplification for the wavelength above 700 nm.  

\begin{figure}
\begin{center}
\includegraphics[scale=0.3]{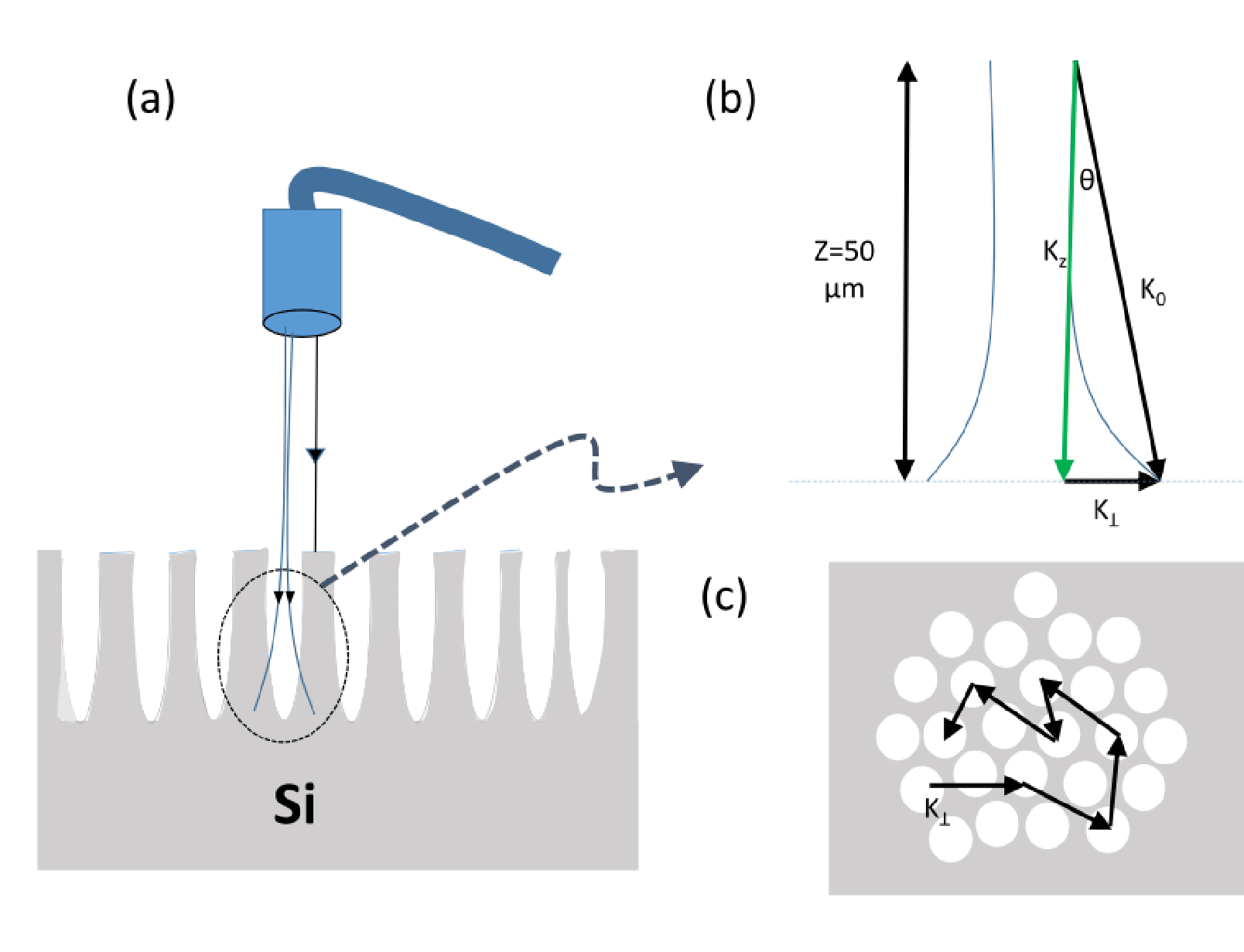}
\end{center}
\vspace{-15pt}
\caption{(Color online) a: Schematic representation of the reflection from PSi and b: spread of the light after multiple scattering from pore walls leading to the localization c: top view of the transverse wave propagation through the pores. $k^{2}_{\bot}=k^{2}_{x}+k^{2}_{y}=k^{2}_{0}-k^{2}_{z}$.}
\label{sch}
\end{figure}

We believe that two different mechanisms are responsible on the reflection (absorption) of the light from (dis)ordered PSi media. The first mechanism is based on the direct absorption of the lights having energies higher than the silicon energygap (1.1eV) inside of the pores due to multiple scattering. The second mechanism is based on the transverse Anderson localization of the lights diffused and subsequently suppressed within the holes which leads to destructive interferences (see Fig.\ref{sch}). It seems that the former mechanism is more relevant to the lower wavelengths, while the later mechanism is more relevant to the higher wavelengths. According to this explanation, presence of a level of ordering can allow photons to escape from the pores and vice versa. This effect is more pronounced at wavelengths comparable with pore diameter or periodicity of the pore arrays. Fig. \ref{sch} visualizes the reflection, scattering and localization processes.   

According to the Anderson theory of light localization, destructive interference of waves can hamper their propagation in a disordered media \cite{sagev2013,walter2008,kroha1993,lee2012,raedt1989,riboli2011}. It is known that there is a threshold value for energy of particles in which below that value the particles are localized. The localization length of these particles is defined by strength of multiple scattering and sample size. Another localization condition is proportionality of scattering length with wave length of electrons or photons ($\lambda \approx$grain size). In addition, we know that the localization in 3D can occur only above some certain value of the disordering. 

However, previous researches have reported only transverse localization of light in transport measurements, but here we show that it is also applicable for reflection experiments. According to this experiment at low frequencies with wavelengths comparable to the pore size(lattice priodicity), there is a tunneling of the electromagnetic waves between pores which leads to the scattering from them and finally localization in disordered pore arrays. We believe this phenomenon plays main role in the absorption of light and hence antireflective behavior of the PSi. Therefore, tuning of the disorder can enable us to engineer light localization and antireflectivity. 

\subsection{Theoretical calcultions}
By considering the PSi as a spatial varying of refraction index in the transverse direction (in plane), it can be assumed that electric field (E$_{\alpha}$, $\alpha=x, y, z$) of traveling waves inside of the pores obey a stochastic Helmholtz equation. The solution of this equation usually provides a time dependent response to the external stimulant in a disordered media (classical approach). In the paraxial limit the envelope of this electric fields obeys the paraxial Schr\"odinger equation which has been frequently used in previous researches (quantum approach). Propagators as a semiclassical objects can connect these two approaches and explain the evolution of the particles (quantum waves where there is no well defined velocity and position) and can solve this equation\cite{sheng1995}.  

Randomly distributed pores plays an important role in spread of the E$_{x,y}$. It is assumed that the fluctuation of the in plane wave number $k^{2}_{\bot}$ in the complex form ($\eta<<1$)are very small in order of 10$^{-9}$. $\Sigma_{k}$ with sums over all impurity momentum is known as a self energy of photons scattered from the pores. This term describes the whole perturbations that is felt by a photon in the presence of the pores. At higher impurity densities (in our case indirectly proportional to the order parameter) in which the scattering events from different impurities begin to interfere, some new physical effects, such as localization can emerge. In the strong scattering regime the self consistent Born approximations (SCBA) can be employed. 

Starting from single particle green function in accordance with Ref.\cite{walter2017} is 
\begin{equation}
\textbf{G}(k^{2}_{\bot}+i\eta)=\frac{1}{-(k^{2}_{\bot}+i\eta)-K^{2}_{0}\sum(k^{2}_{\bot}+i\eta)+q^{2}}
\end{equation}
where q is the transverse wavevector. On the other hand, From two-dimensional version of SCBA equation for the self-energy the green function can be written in the form of
\begin{equation}
\textbf{G}(k^{2}_{\bot}+i\eta)=\frac{2\sum (k^{2}_{\bot}+i\eta)}{K^{2}_{0}\gamma}
\end{equation}

where $\gamma$ is known as disorder parameter. This parameter depends on the relative in-plane change of the dielectric constant. The $\sum(k^{2}_{\bot}+i\eta)$ is self energy of a photon interacting with disordered pores. 
By numerically solution of these two self consistent equations for q=0, the real and imaginary parts of the green function, self energy and waveve vector (related to self energy) can be derived as 

\begin{equation}
G(k^{2}_{\bot}+i\eta)=G^{r}(k^{2}_{\bot}+i\eta)+iG^{i}(k^{2}_{\bot}+i\eta)\\
\end{equation}
\begin{equation}
\sum(k^{2}_{\bot}+i\eta)=\sum^{r}(k^{2}_{\bot}+i\eta)+i\sum^{i}(k^{2}_{\bot}+i\eta)\\
\end{equation}
\begin{equation}
K_{\sum}=K^{r}_{\sum}+iK^{i}_{\sum}\\
\end{equation}

By the obtained real part of the complex self energy as well as imaginary part of the Green's function, diffusion length can be calculated.
\begin{equation}
D_{0}(k^{2}_{\bot})=K^{r}_{\sum}(k^{2}_{\bot})l_{tr}(k^{2}_{\bot})
\end{equation}
where $l_{tr}(k^{2}_{\bot})=\pi K^{r}_{\sum}(k^{2}_{\bot})/G^{i}(k^{2}_{\bot})q^{2}_{c}k^{2}_{0}\sum^{i}(k^{2}_{\bot})$ is called transport mean free path. The $q_{c}$ is a cut-off parameter in which above that the localization can not occur. $K^{r}_{\sum}$ is optical equivalent of lower limit of mean free path or inter atomic spacing in electronic systems with inclusion of disorder effect \cite{ramakrishnan1985}. The imaginary part of the Green function has the meaning of the wave functions decay in which sharp decaying of the wave function can lead to decreasing of the transport mean free path and hence reduction in localization length. The localization length, $\xi(E)$, is exponentially depend on the mean free path and lower limit of mean free path. We employ the method of Ref.\cite{walter2017} by
\begin{figure}
\begin{center}
\includegraphics[scale=0.35]{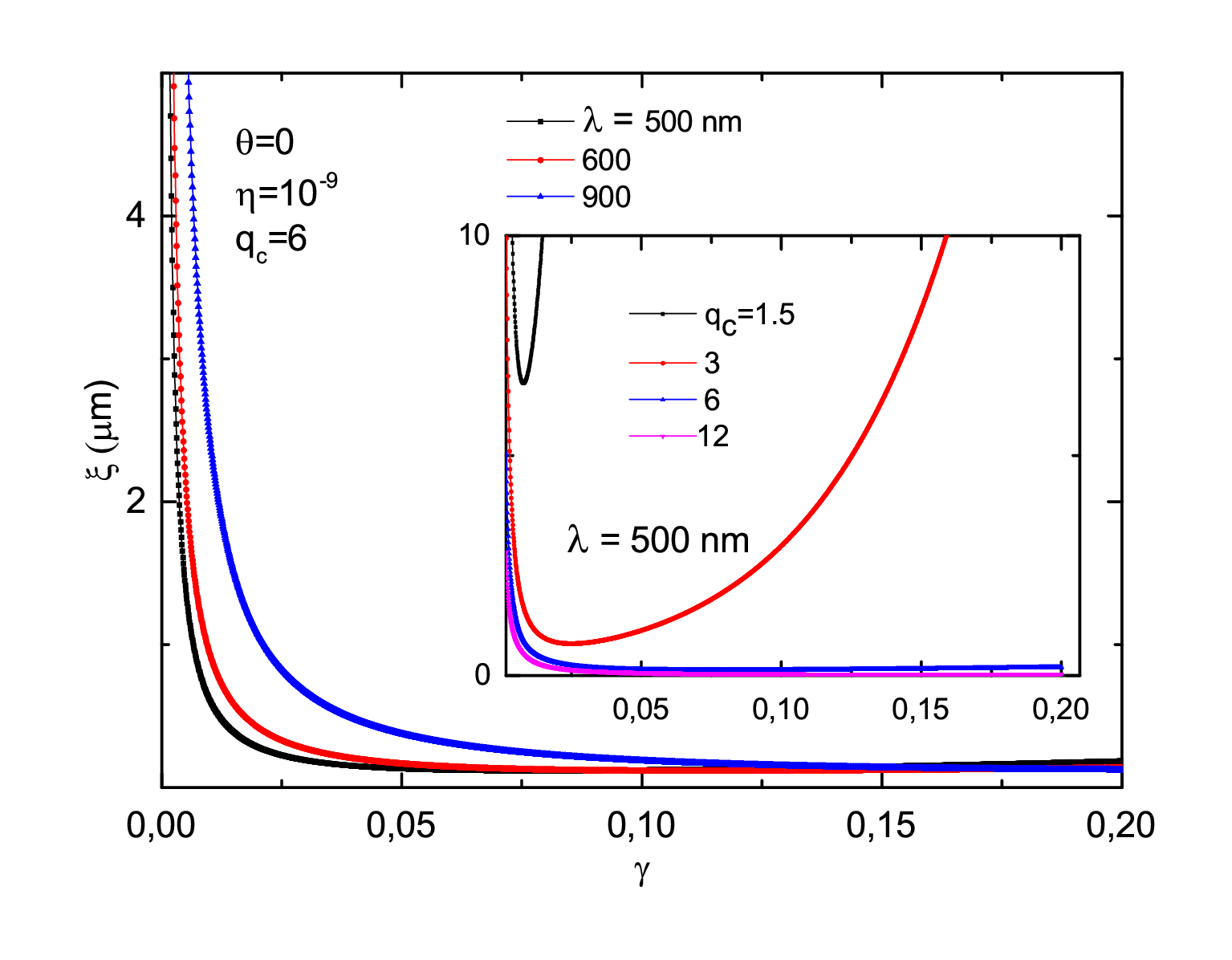}
\end{center}
\vspace{-15pt}
\caption{(Color online) calculated localization length versus disorder parameter for different wavelengths. The inset shows the effect of pore size on the localization length at 500 nm wavelength.}
\label{sim}
\end{figure}

\begin{equation}
\xi(E)=l_{tr}e^{\frac{\pi}{2} D_{0}(k^{2}_{\bot})}
\end{equation}

Our calculations show that for $1\mu$m pore diameter and 20 percent disorderness the $D_{0}$ can change from 260 nm at $\lambda=500$ nm to 480 nm at $\lambda=900$. By increasing the pore size defined by the $q_{c}$, the $D_{0}$ can rise up to $2.4\mu$m.  

Fig.\ref{sim} exhibits numerical calculation of the localization length versus disorder parameter at different wavelength\cite{schwartz2007}. We observed that by increasing the wavelength at some certain order parameters, the localization length increases and therefore the reflection intensity should be stronger. This effect can be observed in very well agreement with our experimental data. Analysis of the inset in this figure indicates that by increasing the pore size (inversely proportional to the $q_{c}$) the localization length is suppressed at higher disorder parameter.
\section{Conclusion}

In conclusion, our experiment in agreement with theory directly visualizes the transverse Anderson localization of light. This research is crucial in the study of transition from extended to localized states in nano(micro)structures. Particularly, we found that the transverse localization of light in randomly distributed pores can increase the probability of the photon absorption which can be important in wide range of applications from energy harvesting to image transport devices and opens a new paradigm for low cost fabrication of optoelectronic devices based on disorder tuning. The PSi as a low cost and high quality template can be employed in different forms, size and morphology to investigate on the other aspects of the light-matter interaction. Detailed mechanism and quantitative explanation of the light absorption in presence of Anderson localization as well as pore characteristics is still to be elaborated. Detailed understanding of such mechanisms and answering to other possible open questions can be addressed in future studies.

\begin{acknowledgments}
S. Erfanifam acknowledges the support from the Iranian Elite's Foundation and valuable consultations of A. H. Baradaran.
\end{acknowledgments}


\end{document}